\documentclass[twocolumn,aps,showpacs,prl]{revtex4-1}
\usepackage{amsmath,amssymb}
\usepackage{subfig}
\usepackage{graphicx}
\usepackage[usenames]{color}
\usepackage{hyperref}
\usepackage{float}

\begin{document}

\title{Many body localization in the presence of a single particle mobility edge}

\author{Ranjan Modak} 
\author{Subroto Mukerjee }
\affiliation{Department of Physics, Indian Institute of Science, Bangalore 560 012, India}

\begin{abstract}
In one dimension, noninteracting particles can undergo a localization-delocalization transition in a quasiperiodic 
 potential. Recent studies have suggested that this transition transforms into a Many-Body Localization (MBL) transition upon 
 the introduction of interactions. It has also been shown that mobility edges can appear in the single particle spectrum for 
 certain types of quasiperiodic potentials. Here we  investigate the effect of interactions in two models with such mobility edges. 
Employing the technique of  exact diagonalization for finite-sized systems, we calculate the level spacing distribution, 
time evolution of entanglement entropy, optical conductivity and return probability  to detect MBL. We find that MBL does indeed occur in one of the two models we study but the entanglement appears to grow faster than logarithmically with time unlike in other MBL systems.  
\end{abstract}


\maketitle

\paragraph*{Introduction:}
 Noninteracting particles in the presence of disorder exhibit the phenomenon of Anderson localization
~\cite{anderson.1958}. 
In  one and two dimensions an arbitrarily weak amount of disorder is sufficient to localize all eigenstates
~\cite{tvr.1979,tvr.1985}. In three dimension a mobility edge, defined as  a threshold eigenstate with energy $E_c$ that 
separates localized and delocalized states can exist. The question of how Anderson localization is modified in the presence of interactions has become  
an area of intense activity following the seminal work of Basko, Aleiner, and Altshuler~\cite{basko.2006}. These authors argued that an 
interacting many-body system can undergo a so called Many-Body Localization(MBL) transition 
in the presence of quenched disorder. This MBL transition involves highly excited many-body quantum states and can thus extend up to even infinite temperature in contrast to a usual quantum phase transition~\cite{sachdev.2007}, which involves only the 
ground state. Traditional notions of statistical mechanics do not apply to this transition and the localized phase, including the Eigenstate Thermalization Hypothesis (ETH)~\cite{deutsch1991,srednicki1994,rigol.2008} for the mechanism of thermalization in isolated quantum systems~\cite{vadim.2007,pal.2010}.  It has thus 
been suggested that there are emergent conservation laws for these localized systems~\cite{huse2014phenomenology,modak.2015} like for integrable ones, which too do not thermalize~\cite{rigol1.2009,santos.2010}.

It is possible to have a localization-delocalization transition similar to the MBL transition for non-interacting one dimensional models with quasi-periodic potentials. 
An example of such a model is the Aubry-Andre model~\cite{aubry.1980}(AA model) , which has the form
\begin{equation}
H=\sum_{i}h_{i}n_{i} -t\left( c^{\dag}_{i}c_{i+1}+c^{\dag}_{i+1}c_{i} \right),
s\label{Eq:AAmodel}
\end{equation}
where $c$ ($c^\dagger$) annihilates (creates) spinless fermions. $t$ is the hopping and $h_i$ an onsite potential 
with the quasi-periodic form $h_{i}=h\cos(2\pi \alpha i +\phi)$, where $\alpha$ is an irrational number and $\phi$ an offset. This model has a localization-delocalization transition at $h=2t$, where {\em all} states are (de)localized for $h (<) > 2t$. A numerical 
study of this model with a nearest neighbor interaction of the form $V\sum_in_i n_{i+1}$ has shown that the single particle transition changes into an MBL transition 
akin to the one in models with on-site disorder~\cite{huse.2013}. Furthermore, this model has recently been emulated in experiments on 
cold-atoms in the non-interacting limit~\cite{fallani.2007,lucioni.2011} and with interactions to observe MBL~\cite{ehud.2015}.

Modifications to the AA model have been proposed to yield models which possess single-particle mobility edges~\cite{griniasty.1988,dassarma.1990,ganeshan.2014}. It has been argued that in the presence of (even weak) interactions, localized single particle states can thermalize when coupled to a bath of even a single delocalized state that is protected topologically or otherwise~\cite{nandkishore2014marginal}.  However, a very recent work shows that in the absence of any such protection, under certain conditions, the localized states can instead thermalize the bath ~\cite{nandkishore2015many}. Models with single particle mobility edges are ideal to study the latter scenario since the delocalized states have no protection from localization.

In this work, we study whether MBL occurs in models with single particle mobility edges upon switching on weak to moderately strong interactions. Employing exact diagonalization of finite-sized systems, we calculate  various diagnostics to detect MBL such as the level spacing distribution, time evolution of entanglement entropy, optical conductivity and return probability.  Our conclusion is that MBL occurs in one of the models we study but not the other. However in the localized phase we observe, the entanglement entropy increases appears to increase linearly with time (like in an ergodic phase) but saturates to a sub-thermal value characteristic of MBL. The growth of entanglement entropy with time in a regular many-body localized phase is logarithmic~\cite{bardarson.2012,vosk.2013,prosen.2008}. All other diagnostics appear to be consistent with regular MBL. We examine the possible reasons for the different behaviors of the two models and also provide possible reasons for the observed linear growth of entanglement with time.

We have studied two different interacting one-dimensional models of spinless fermions, which in the non-interacting limit have single particle mobility edges.  The first, which we shall refer to as model I is described by the Hamiltonian
\begin{equation}
  H=\sum_{i}h_{i}n_{i} -t(c^{\dag}_{i}c_{i+1}+c^{\dag}_{i+1}c_{i})+Vn_{i}n_{i+1}
  \label{Eq:modelsham}
\end{equation}
where $h_{i}=h\cos(2\pi \alpha i^{n} +\phi)$  with $0<n<1$. For $V=0$ and $n=1$, this is just the AA model. However, for $n<1$ and $V=0$, the model has a single-particle
mobility edge when $h<2t$~\cite{griniasty.1988,dassarma.1990}. All single particle states with energy between $\pm |2t-h|$ are delocalized and 
all other states are localized. For $h>2t$ all single particle states are localized as in the usual AA model.

The other model (which we refer to as model II) is also of the form in Eqn.~\ref{Eq:modelsham} but with 
$$ h_{i}=h\frac{1-\cos(2\pi i \alpha +\phi)}{1+\beta \cos(2\pi i \alpha +\phi) },$$ 
with $\beta \in (-1,1)$. When $\beta=0$ and $V=0$, this model also reduces to the AA model. 
For $V=0$, there is a mobility edge separating, localized and extended states at an energy $E$ given by
$\beta E=2(t-h/2)$. This model also can be experimentally realized~\cite{ganeshan.2014}.

We have studied both models using exact diagonalization on finite-sized systems up to size $L=16$ (data in plots shown for $L=14$) with open boundaries and have averaged 
 over the offset $\phi$ for better statistics. $t=1$ and $\alpha=\frac{\sqrt{5}-1}{2}$ in all our calculations. We now discuss our results.
 \begin{figure}[H]
\includegraphics[height=3.0 in,width=1.8in,angle=-90]{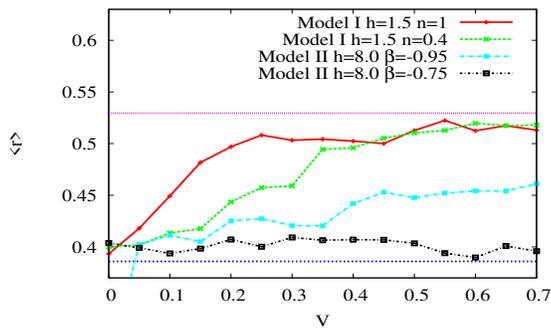}
\caption{(Color Online)The variation of the mean of the ratio between adjacent gaps in the spectrum for $L=14$ at half filling
for model I and model II. The blue dotted line is for the Poissonian distribution and the pink one is
for the Wigner-Dyson distribution.}
\label{Fig:level spacing}
\end{figure}
 \paragraph*{Energy level spacing statistics:}
Energy level spacing statistics is often used to characterize the MBL transition. 
There is a crossover from a Wigner-Dyson to Poissonian distribution upon going from the ergodic to many-body localized phase, which can be tracked by the ratio of successive gaps, $r_{n}=\frac{min(\delta_n,\delta_{n+1})}{max(\delta_n,\delta_{n+1})}$~\cite{vadim.2007},
%\begin{equation}
 %r_{n}=\frac{min(\delta_n,\delta_{n+1})}{max(\delta_n,\delta_{n+1})},
%\end{equation}
where $\delta_n=E_{n+1}-E_{n}$, the difference in energy between the $n^{\rm th}$ and $n+1^{\rm st}$ energy eigenvalues. For a Poissonian (Wigner-Dyson, specifically of the Gaussian Orthogonal type) distribution, the mean value of $r$ is $2\ln 2-1\approx 0.386$ ($\approx 0.5295$). The distribution function $P(r) \rightarrow 0$, as $r \rightarrow 0$ in the presence of level repulsion.

For model I, with $V=0$ , $h<2$ and $n=1$, all single particle states are delocalized. As $V$ is increased, the level 
spacing distribution starts to follow the Wigner-Dyson distribution. For, $n<1$, with a mobility edge, level statistics 
obey the Wigner-Dyson distribution, even though there are localized states as shown in Fig.~\ref{Fig:level spacing}.
Deep in the localized phase ($h>>2$), increasing $V$ yields a Poissonian distribution in both cases ($n=1.0$ and $n<1.0$) 

Unlike for model I, the position of the mobility edge in the non-interacting limit of model II can be tuned by varying the parameters $\beta$ and $h$~\cite{ganeshan.2014}. We choose, $h=8$ and change $\beta$ from -0.95 to 0 so the fraction of single particle localized states increases progressively. In contrast to model I, here the level spacing distribution appears to be Poissonian for $V \neq 0$ as can be seen in Fig.~\ref{Fig:level spacing}.

\begin{figure}
\includegraphics[height=3.0 in,width=1.8in,angle=-90]{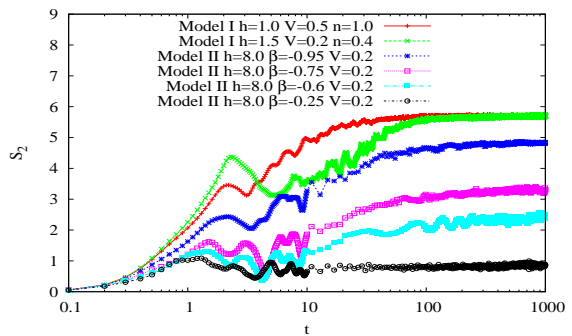}
\caption{(Color Online)Variation of the Renyi entropy for $L=14$ at half filling for the two models with different parameters.}
\label{Fig:entropy II}
\end{figure}

\begin{figure}
\begin{center}
\setlength{\unitlength}{8.5cm}
\begin{picture}(1.5, 0.818)(0,0)
   \put(0,0.8){\resizebox{1\unitlength}{!}{\includegraphics[angle=-90]{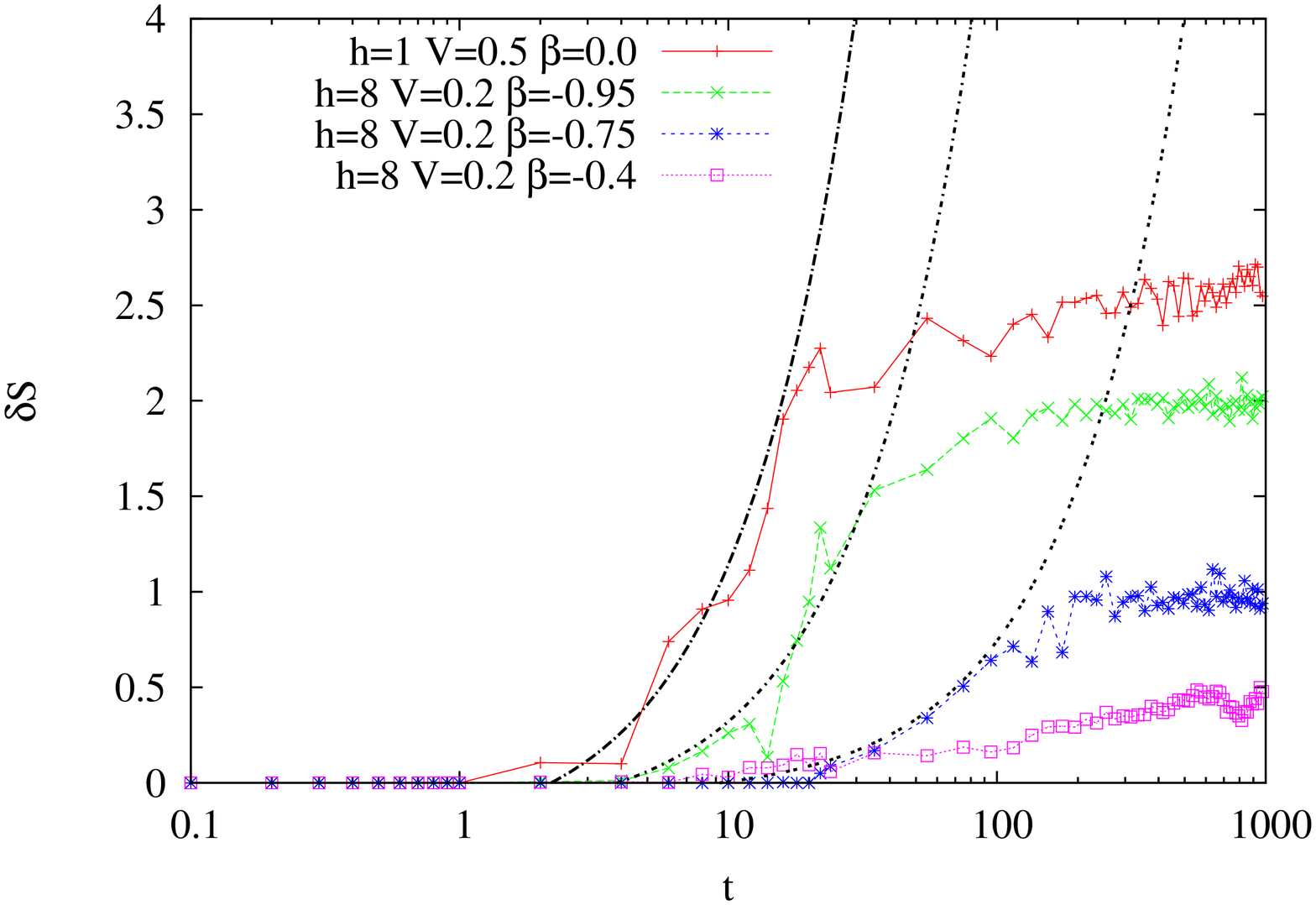}}}
\put(0.15,0.63){\resizebox{0.41\unitlength}{!}{\includegraphics[angle=-90]{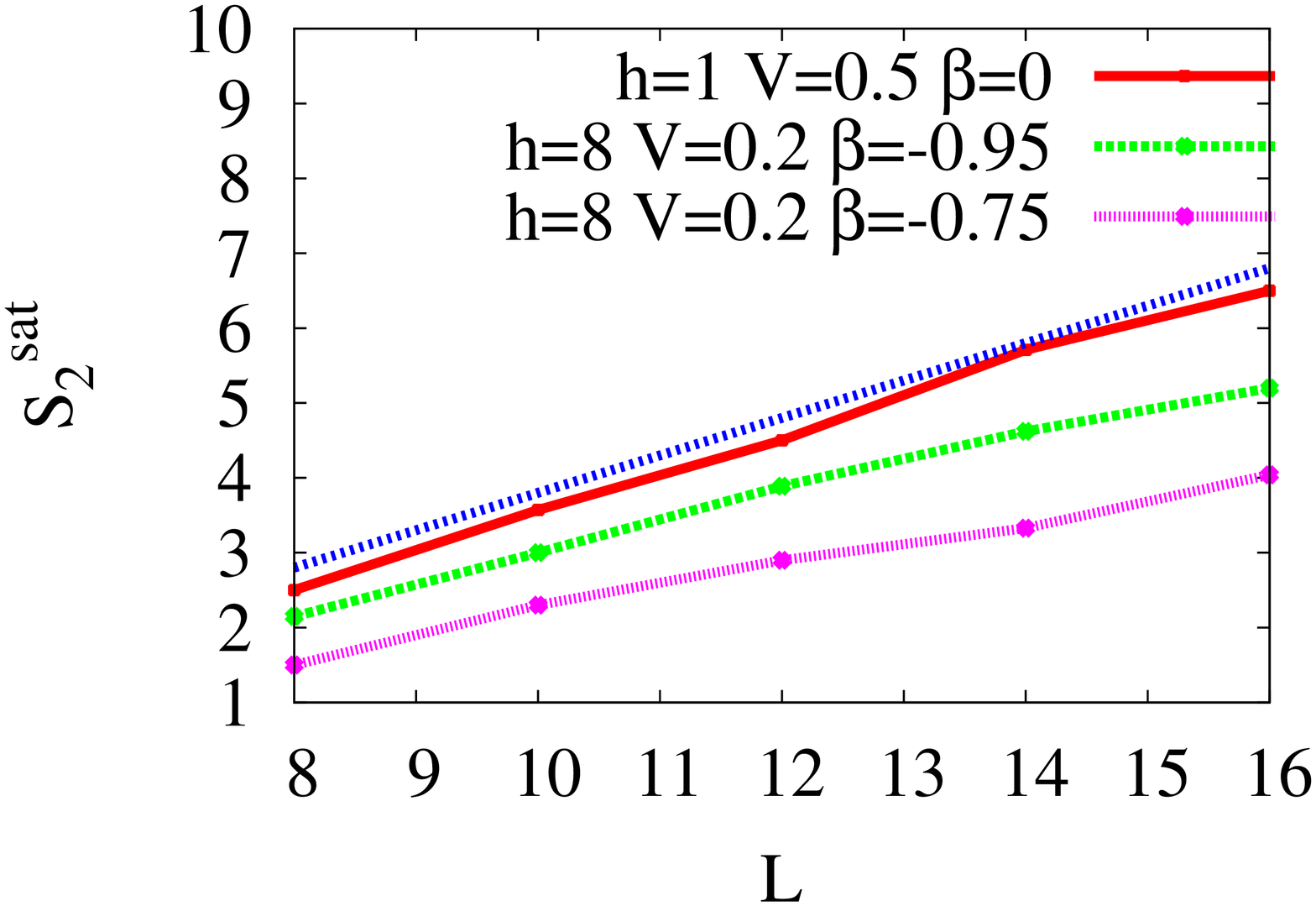}}}
\end{picture}
\end{center}
\vspace{-0.6in}
\caption{(Color Online)The variation of $\delta S=S_2(t,V)-S_2(t,V=0)$. $S_2$ is the Renyi entropy for $L=14$ at half filling for 
the two models with different parameters. The dotted lines are linear fits in $t$. 
(Inset) The variation of the saturation value of $S_2$ with $L$. The blue dotted line corresponds to 
thermal value of $S_2=\frac{L}{2}-1.2$ for system size $L$.}
\label{Fig:delta S}
\end{figure}

\paragraph*{Entanglement entropy:}
The entanglement entropy is another diagnostic that can be used to distinguish between the ergodic and many-body localized phases.
We have studied the time evolution of the entropy $S(t)$ by sampling the initial unentangled states at random over the entire energy spectrum, which is equivalent to working 
at infinite temperature~\cite{huse.2013}.
$S(t)$ has been argued to grow linearly in the ergodic phase and logarithmically in the many-body localized
phase~\cite{bardarson.2012,serbyn.2013}. 

The system of length $L$ is divided into two equal parts $A$ and $B$. Our calculation is of the order 2 Renyi entropy $S_2(t)=-\log_2(Tr_{A}{{\rho_{A}(t)}^{2}})$ (which is computationally less expensive than the von-Neumann entropy) ~\cite{rrnyi1961measures} , where $\rho_A(t)$ is the reduced density matrix of $A$ obtained from the instantaneous state of the full system. It is known that in the ergodic phase, $S_{2}(t)\sim t$ at long times and saturates to the infinite temperature thermal value while for the usual many-body localized phase with weak interactions, $S_2(t)\sim \zeta\log(t)$, where $\zeta$ is the localization length of the single particle eigenstates. It saturates to a value much smaller than the thermal value, but which is still extensive in system size.  For our system, the infinite temperature $S_2\sim\frac{L}{2}-1.2$ for system size $L$~\cite{huse.2013}.

For model I with a single-particle mobility edge, $S_2(t)$ increases linearly with time but then appears to saturate to the thermal value as shown in Fig.\ref{Fig:entropy II}. However, for model II, $S_2$ appears to grow linearly with time but saturates to a value smaller than the thermal value. This can be seen from Fig.~\ref{Fig:entropy II}, where the time evolution of  $S_{2}$ has been plotted for model II for  $V=0.2$ , $h=8$  and $\beta= -0.95$ ,$-0.75$ and $-0.6$, with progressively increasing fractions of single-particle localized states. The saturation value depends on the number of localized single particle states: As the fraction of single-particle delocalized states  increases, so does the saturation value.

To confirm the linear growth $S(t)$, we have plotted $\delta S=S_2(t,V)-S_2(t,V=0)$
in Fig.~\ref{Fig:delta S} as a function of time. At very early times 
$S_2(t,V)$ and $S_2(t,V=0)$ tend to coincide, reflecting the formation of short range  entanglement at the 
cut between the subsystems. Then, $S_2(t,V=0)$ saturates but for the interacting system, 
$S_2$ keeps growing with time as shown in Fig.~\ref{Fig:delta S}. At intermediate times, 
as long as there is a mobility edge in the single particle spectrum, 
$\delta S$ fits quite well to a linear function of time. When all single particle states are localized, 
the growth of $\delta S$ as a function of $t$ is much slower than linear and possibly logarithmic. At long times, $\delta S$ 
saturates to a sub-thermal value in all cases (For a calculation to even longer times see the supplementary information~\cite{suppl}).

We have also plotted the saturation value of $S_{2}$ as a function of system size $L$.  As shown in the inset of Fig.~\ref{Fig:delta S}
$S_{2}^{sat} \sim L$ for the ergodic phase as well as for the model with a mobility edge. This plot also shows that 
the $S_2^{sat}$ curve for the system with the single particle mobility edge system does not intersect the curve for the ergodic 
system when extrapolated to the thermodynamic limit. Thus, the saturation of the entropy to a sub-thermal value is not a finite-size effect.
%\begin{figure}
%\includegraphics[width=2.5in,angle=-90]{sigma_diff_model.eps}
%\caption{Variation of $\sigma(\omega)$ with $\omega$ for different models for L=14 at half filling.
%Dashed lines are best fitted lines.}
%\label{Fig:conductivity}
%\end{figure}
\begin{figure}
\begin{tabular}{cc}
\includegraphics[height=1.8in, width=1.5in,angle=-90]{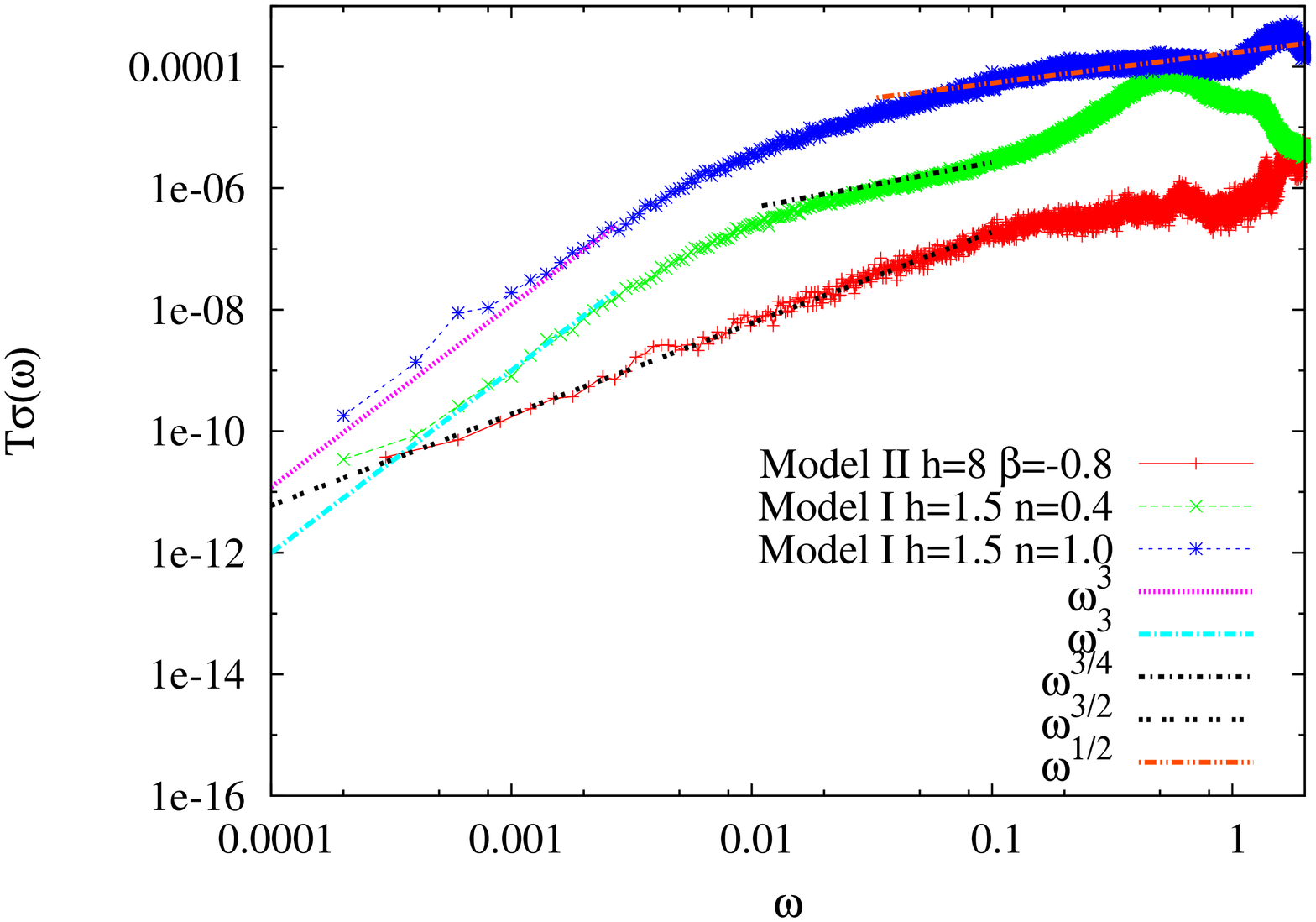} &
\includegraphics[height=1.8in, width=1.5in,angle=-90]{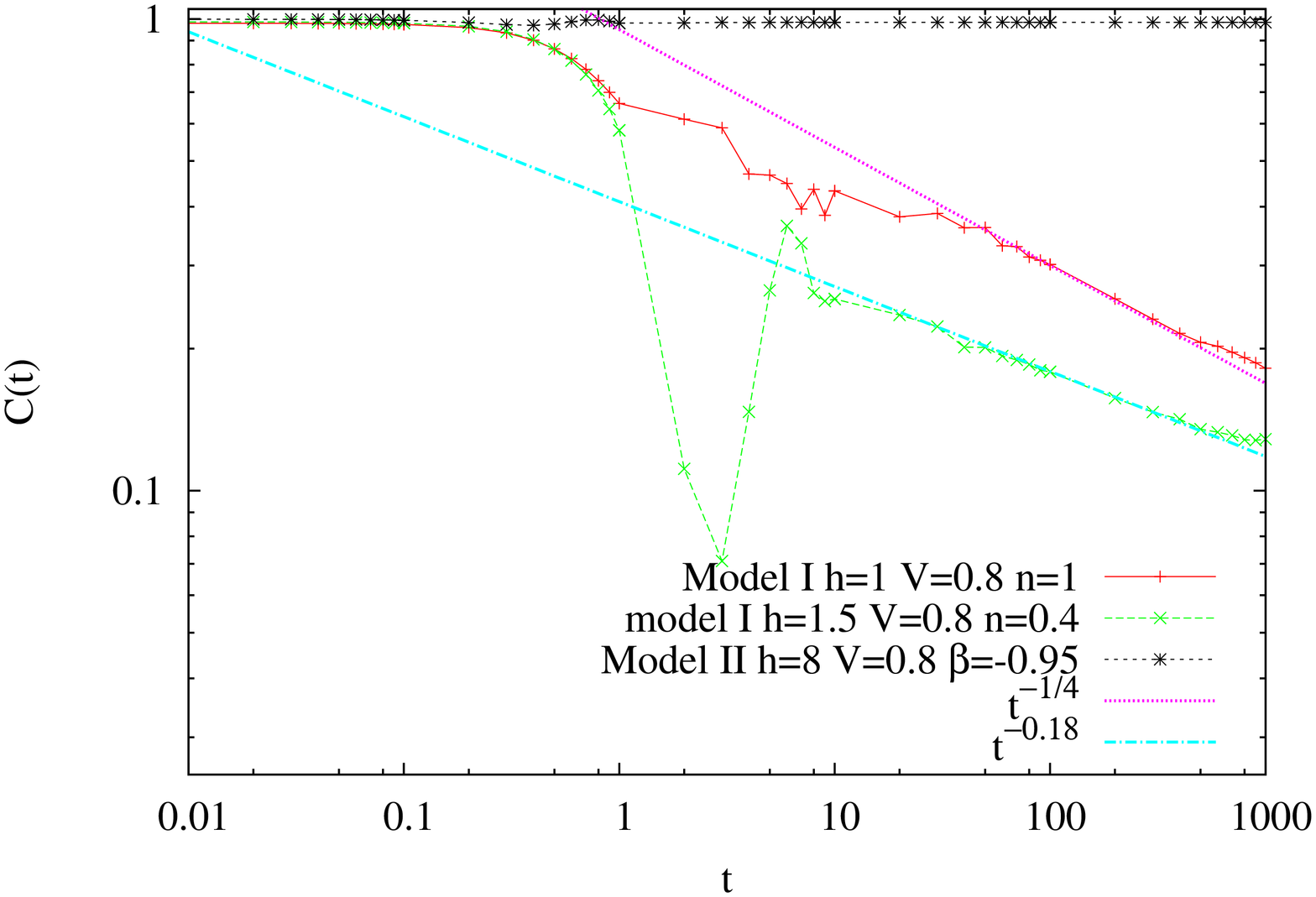}
\end{tabular}
\caption{(Color Online)(A)The variation of $\sigma(\omega)$ with $\omega$ for the two models for $L=14$ at half filling for $V=0.8$.
The rescaled values of $\sigma$ are plotted . $10\sigma$, $\sigma$ and $0.1\sigma$ are plotted respectively for 
model I ($n=1$, $h=1.5$ and $n=0.4$, $h=1.5$) and model II ($h=8$, $\beta=-0.8$). The dashed lines are the best fit lines.
(B)The variation of the return probability $C(t)$ as a function $t$ for different models for $L=14$ at half filling.
The dashed lines are the best fit lines}
\label{Fig:conductivity2}

\end{figure}

\paragraph*{Optical conductivity:}
The optical conductivity $\sigma(\omega)$ is another diagnostic that can be used to identify the ergodic and many-body localized phases. 
In the case of a clean metal, the DC conductivity $\sigma(\omega=0)\neq 0$ with a frequency-dependent additive term that goes as $\omega^{1/2}$ at high 
temperature~\cite{subroto.2006}. In the presence of disorder, a subdiffusive phase  can exist even on the thermal side of 
the MBL transition~\cite{vosk2014theory,lev2014absence,agarwal.2014, potter2015theory}, for which $\sigma(\omega)\sim\omega^{a}$ with $0<a<1$. 
In the many-body localized phase, $\sigma(\omega)\sim \omega^{a} $ with $1\leq a < 2$ ~\cite{gopalakrishnan.2015} and
 $a\to 1$ as the transition is approached. $\sigma(\omega)$ is given by the Kubo formula,
\begin{equation}
 T\sigma(\omega)=\frac{1}{ZL}\sum_{mn}
 |<m\sum_{i}j_i|n>|^{2}\delta(\omega-E_m+E_n) 
 \end{equation}
as $T \rightarrow \infty$, where, $m$,$n$ are the many body eigenstates of the system with energies $E_m$ and $E_n$. $j_i$ is the local current density. 

As shown in Fig.~\ref{Fig:conductivity2}(A) for model I at very low values of $\omega$, $\sigma(\omega)\sim \omega ^{3}$.
This is from a combination of level repulsion (one power of $\omega$) and 
open boundary conditions ($\omega ^{2}$)~\cite{gopalakrishnan.2015}. Subtracting this out, we obtain
$\sigma \sim \omega ^{1/2}$ and $\sigma \sim \omega^{3/4}$ for model I with $h=1.5$, $n=1.0$ and $h=1.5$ and $n=0.4$ respectively. 
For model II in the presence of a single-particle mobility edge, after subtracting out the $\omega^2$ 
dependence~\footnote{Since there is no level repulsion in this model, the factor that has to be subtracted goes as $\omega^2$ and not $\omega^3$.}, $\sigma(\omega)\sim \omega ^{a}$ at low frequencies with $1\leq a<2$ like in the usual many-body localized phase. In Fig.~\ref{Fig:conductivity2}(A) for a particular choice of parameter $\beta$, $\sigma \sim \omega ^{3/2}$. 

We have verified that we obtain the same exponent $a$ even with periodic boundary conditions, where the subtraction is of a different power of $\omega$. Further, we find that the exponent for model II increases as the fraction of localized states for $V=0$ increases consistent with the expectation that the system gets pushed deeper into the many-body localized phase if it starts with more localized states without interactions.
 
 %\begin{figure}
%\includegraphics[height=3.0in, width=1.8in,angle=-90]{corr.eps}
%\caption{(Color Online)The variation of the return probability $C(t)$ as a function $t$ for different models for $L=14$ at half filling.
%The dashed lines are the best fit lines.}
%\label{Fig:return}
%\end{figure}

\paragraph*{Return probability:}
 The return probability $C(t)$, measures the probability of particles to
 return to their initial positions during the evolution of the system and is defined as 
\begin{equation}
 C^{j}(t)=\frac{4}{Z}\sum_{nm}e^{-i\omega_{mn}t}|<n|(n_j -1/2)|m>|^2
\end{equation}
 where, $Z$ is the Hilbert space dimension.
 We have calculated $C(t)=\frac{1}{L}\sum_{j}C^{j}(t)$ with $C(t=0)=1$. In the ergodic (diffusive) phase, $C(t)\sim t^{-1/2}$ and in the many-body localized phase,
 it remains finite in the long time limit ~\cite{agarwal.2014}. The behaviour of $C(t)$ at long times is drastically different for the two models as can be seen in Fig.~\ref{Fig:conductivity2}(B).  For model I, with $n=0.4$ and $h=1.5$  at long times, $C(t)\sim t^{-b}$ with $b=0.18$ and for 
model II with $\beta=-0.95$ and $h=8$, $C(t)$ does not decay with time. The result for model I is consistent with the scaling relation $a+2b=1$, proposed  
by Agarwal et al.~\cite{agarwal.2014}. We note however that we have not been able to clearly observe $C(t)\sim t^{-1/2}$  in the thermal phase probably due to limitations of system 
size.

\paragraph*{Discussion:}
We have demonstrated the effect of interactions on models with mobility edges in the non-interacting limit.
Our numerical results employing a number of different diagnostics show that an MBL phase can occur 
in such a situation. We find that model II displays MBL while model I does not. One possible reason for this is finite size effects.  To examine this, we have calculated the Inverse Participation Ratio (IPR) for all the states of an $L=14$ system with $V=0$

\begin{figure}
\includegraphics[height=3.0 in ,width=1.8in,angle=-90]{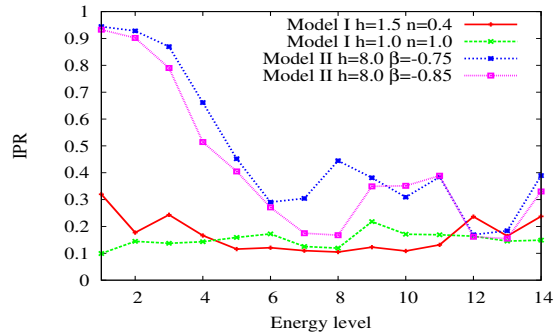} 
\caption{(Color Online) IPR as a function of energy levels for the two models for $L=14$.}
\label{Fig:IPR}
\end{figure} 

The IPR of a normalized eigenstate $\psi$ is defined $IPR_{\Psi }=\sum_{j}|c_j|^{4}$, where  $c_j$ is the amplitude of $\psi$ at site $j$. $IPR  \sim 1$ for a localized state and is much smaller (typical $ \sim 1/L$) for a delocalized one. The IPR values for the two models are shown in Fig.~\ref{Fig:IPR}. It can be seen that while there are localized states (with IPR of order 1) along with delocalized ones for model II, the states of model I appear to be delocalized for our system size. This behavior presumably persists even with interactions (which generally tend to cause delocalization). As a result, none of the diagnostics for this model show any evidence of many-body localization, to observe which probably requires larger system sizes that are not easily accessible with exact diagonalization. Another possibility is that even in the thermodynamic limit  the localized states of Model I are only ``weakly'' localized compared to those of Model II and thus fail to localize the bath. Another possibility is that the delocalized states of model I are ``inherently'' more robust compared to the localized ones whereas for model II, it is the other way around. A calculation of the IPR for large system sizes seems to suggest that this is true~\cite{suppl}. Additional calculations of the matrix elements for flip-flop processes and the localization length exponent $\nu$\cite{supple} seem to indicate that model II may not satisfy the conditions for delocalization even when coupled to a protected bath~\cite{nandkishore2014marginal}. Thus, an introduction of interactions would tend to cause MBL in model II and thermalization in model I even in the thermodynamic limit. Additional studies are required to fully understand the differences between the two models.

For model II, the entanglement entropy appears to grow linearly with time (instead of logarithmically) before saturating to a sub-thermal value. A possible explanation is the simultaneous but independent contributions of the delocalized and localized states which individually would produce linear and logarithmic growth respectively. For sufficiently long times, 
the linear growth would dominate, which is what we observe. A mechanism has been proposed recently invoking the idea of rare thermal regions in a many-body localized phase~\cite{gopalakrishnan.2015} mainly to explain the  behavior of $\sigma(\omega)$ near the MBL transition. The specific systems studied has spatially separated ergodic and thermal regions. A calculation of $S(t)$ performed by us for a similar system yields faster than logarithmic (algebraic) growth of $S_2$ with time, similar to what we observe for model II.
Thus, the delocalized states in our models could be performing a role analogous to that of rare thermal regions and producing a linear growth of entanglement. Note however,  that the quasi-periodic potential in our models is correlated at different sites and so no true rare regions in the sense of~\cite{gopalakrishnan.2015} can actually occur. The analogy is therefore not a deep one. Algebraic growth of entanglement with time in a many-body localized phase has also been observed in the presence of long-range interactions~\cite{pino.2014}. Another possibility is that the apparent linear growth is a finite-size effect in model II and will eventually become logarithmic for sufficiently large system sizes. The slow growth of the localization length near the mobility edge for model II as characterized by the exponent $\nu$~\cite{suppl} might be a sign that finite-size effects are important.

\paragraph*{Note added:} A related study of many-body localization in model II appeared at the same time as ours~\cite{li2015energy} .
\paragraph*{Acknowledgments:} We thank Ehud Altman, David Huse, Kartiek Agarwal, Diptiman Sen and especially Rahul Nandkishore for discussions. RM acknowledges support from the UGC-BSR Fellowship and SM from the DST, Govt. of India and the UGC-ISF Indo-Israeli joint research program for funding.

\end{document}